Crystallographic features of the approximant H ($Mn_7Si_2V$) phase in the Mn-Si-V alloy system

Kei Nakayama [a, *], Takumi Komatsuzaki [a], Yasumasa Koyama [a, b]

[a]Department of Electronic and Physical Systems, Waseda University, 3-4-1, Okubo, Shinjuku-ku, Tokyo 169-8555, Japan

[b]Kagami Memorial Research Institute for Materials Science and Technology, Waseda University, 2-8-26, Nishiwaseda, Shinjuku-ku, Tokyo 169-0051, Japan

*Corresponding author
*Email address:* d.i.y.999@fuji.waseda.jp
*Postal address:* Kagami Memorial Research Institute for Materials Science and Technology, Waseda University, 2-8-26, Nishiwaseda, Shinjuku-ku, Tokyo 169-0051, Japan
*Telephone number:* +81-3-5286-3789
*Fax number:* +81-3-5286-3771

Abstract

The intermetallic compound H ($Mn_7Si_2V$) phase in the Mn-Si-V alloy system can be regarded as an approximant phase of the dodecagonal quasicrystal as one of the two-dimensional quasicrystals. To understand the features of the approximant H phase, in this study, the crystallographic features of both the H phase and the ($\sigma \to H$) reaction in Mn-Si-V alloy samples were investigated, mainly by transmission electron microscopy. It was found that, in the H phase, there were characteristic structural disorders with respect to an array of a dodecagonal structural unit consisting of 19 dodecagonal atomic columns. Concretely, penetrated structural units consisting of two dodecagonal structural units were presumed to be typical of such disorders. An interesting feature of the ($\sigma \to H$) reaction was that regions with a rectangular arrangement of penetrated structural units (RAPU) first appeared in the $\sigma$ matrix as the initial state, and H regions were then nucleated in contact with RAPU regions. The subsequent conversion of RAPU regions into H regions eventually resulted in the formation of the approximant H state as the final state. Furthermore, atomic positions in both the H structure and the dodecagonal quasicrystal were examined using a simple plane-wave model with twelve plane waves.

1. Introduction

The dodecagonal quasicrystal with twelvefold symmetry is one of the two-dimensional quasicrystals having a translational symmetry along one crystallographic direction and has so far been found in macromolecular polymers and colloids as well as in alloys such as Mn-Si-V and Cr-Ni-Si systems [1–11]. Its striking feature is that it has both a twelvefold rotational axis and quasiperiodicity, which can be expressed by a different tiling model from that for the decagonal quasicrystal [12–16]. In the case of the dodecagonal quasicrystal, the tiling model is composed of three kinds of tiles: that is, square-, triangle-, and rhombus-shaped tiles. It is also known that the atomic arrangement in the dodecagonal quasicrystal involves a dodecagonal structural unit consisting of 19 dodecagonal atomic columns, and that a dodecagonal column is formed by connecting complex polyhedra with a coordination number (CN) 14 along a column axis in the crystallographic direction [5,17]. In addition, it has been reported that there are approximant phases having local structural units similar to the dodecagonal quasicrystal [17–19]. The intermetallic compound H ($Mn_7Si_2V$) phase in the Mn-Si-V alloy system is a typical example of an approximant phase of the dodecagonal quasicrystal in alloys [18]. In this study, we focused on the approximant H phase to obtain a deeper understanding of the dodecagonal quasicrystal in alloys.

According to the previously reported phase diagram of the Mn-Si-V alloy system, the approximant H phase with hexagonal P6/mmm symmetry is known to be stable below about 1100 K around the Mn-20at.%Si-10at.%V composition, while the σ phase with tetragonal $P4_2$/mnm symmetry is present, for instance, at 1273 K in a wide composition range around Mn-10at.%Si-15at.%V [20]. Based on this, the (σ → H) reaction can be expected to occur around a composition of Mn-17at.%Si-9at.%V by keeping samples with a metastable σ state below about 1100 K. A notable feature of the crystal structures of the σ and H phases is that they are commonly identified as coordination-polyhedra structures with higher coordination numbers (CNs) such as CN12 and CN14 polyhedra [18,21]. Their crystallographic data reported previously indicate that the two structures involve three kinds of polyhedra; that is, CN12, CN14, and CN15 polyhedra. The important points of the two structures are that the CN14 polyhedron appears with the highest relative frequency among the three, and that they have a dodecagonal atomic column. Note that the dodecagonal atomic column is characterized by a one-dimensional array of CN14 polyhedra. In addition, the dodecagonal structural unit consisting of 19 dodecagonal atomic columns is found only in the H structure, just as in the case of the dodecagonal quasicrystal. The presence of the dodecagonal structural unit is apparently an experimental indication that the H phase in the Mn-Si-V alloy system can be regarded as an approximant phase of the dodecagonal quasicrystal in alloys.

The intermetallic compound H phase in the Mn-Si-V alloy system can be identified as an approximant phase of the dodecagonal quasicrystal. We have investigated the crystallographic features of both the H phase and the (σ → H) reaction in Mn-Si-V alloy samples, mainly by transmission electron microscopy, in order to understand the detailed features of the dodecagonal quasicrystal in alloys. Concretely, two kinds of alloy samples with different thermal treatments were prepared for our present study. One having the Mn-20at.%Si-10at.%V composition was kept at 1073 K to get a single H state as the approximant phase. On the other hand, the (σ → H) reaction was examined by using the other Mn-17at.%Si-9at.%V samples, which were first kept at 1273 K to get the σ phase, and subsequently at 1073 K for the reaction. The latter samples in particular were used to investigate the formation of the dodecagonal structural unit consisting of 19 dodecagonal atomic columns, which occurs in both the approximant H structure and the dodecagonal quasicrystal, but not in the σ structure. We also examined the atomic arrangements of both the H structure and the dodecagonal quasicrystal using a simple plane-wave model having twelve plane waves.

2. Experimental procedure

Two kinds of Mn-Si-V alloy samples with different compositions and different thermal treatments were prepared to examine the crystallographic features of both the approximant H phase and the (σ → H) reaction. Based on the previously reported phase diagram, the nominal compositions of Mn-20at.%Si-10at.%V and Mn-17at.%Si-9at.%V were adopted for investigating the features of the H structure and the reaction, respectively. As for sample preparation, ingots of Mn-Si-V alloys with the noted compositions were first made from metals Mn, Si, and V with purity of 99.9% by an induction-melting technique. Ingots prepared with the Mn-20at.%Si-10at.%V composition for the former purpose were kept at 1073 K for 100 h in vacuum to obtain the single H phase, followed by quenching in ice water. For the latter purpose, ingots with the Mn-17at.%Si-9at.%V composition were first kept at 1273 K for 24 h in vacuum to obtain the single σ phase. Plate-shaped samples with a metastable σ phase were cut from the ingots and annealed at 1073 K in vacuum for various annealing times to induce the reaction. The appearance of metastable and equilibrium states in the samples was checked by measuring their x-ray powder diffraction profiles at room temperature in the angular range of $10° \leq 2\theta \leq 100°$, using a Rigaku-SmartLab diffractometer with Cu$K\alpha$ radiation. To examine the crystallographic features of the samples, we took electron diffraction patterns, bright- and dark-field images, and high-resolution electron micrographs at room temperature with a JEM-3010-type transmission electron microscope with an accelerating voltage of 300 kV.  Thin specimens for observation by transmission electron microscopy were prepared by an Ar-ion thinning technique.

3. Experimental results
A. Mn-20at.%Si-10at.%V alloy samples

Of the two kinds of Mn-Si-V alloy samples, samples with the Mn-20at.%Si-10at.%V composition were used to examine the crystallographic features of the approximant H phase.  Figure 1 shows an x-ray powder diffraction profile of a Mn-20at.%Si-10at.%V sample, which was measured at room temperature in the angular range of $35° \leq 2\theta \leq 55°$.  Stronger reflections are observed around $2\theta = 39°$, $43°$, $45°$, $47°$, and $50°$. A simple analysis of these reflections confirmed that the profile was basically consistent with the H structure with hexagonal P6/mmm symmetry.  A notable feature of the profile is that, in addition to the H reflections, additional reflections with weak intensities are also seen, as indicated by the black arrows, together with a relatively strong background.  This implies that structural disorders probably occurred in the Mn-20at.%Si-10at.%V sample.  To understand the features of the structural disorders, we observed Mn-20at.%Si-10at.%V samples by transmission electron microscopy.  It was found that there were two types of areas in the samples, which are referred to here as Areas I and II, and their detailed features are described below.

Two bright-field images taken from Areas I and II at room temperature are shown in Fig. 2, together with the corresponding electron diffraction patterns in the insets.  The electron beam incidence for these two images is parallel to the $[0001]_H$ direction, where the subscript H denotes the hexagonal system.  Although sharp, dark contrast lines due to line or planar defects are seen in Fig. 2(a) for Area I, a uniform contrast is basically observed in the image.  From an analysis of the electron diffraction patterns taken from Area I, including the pattern in the inset, the state was identified as the hexagonal P6/mmm H state.  That is, the $[0001]_H$ direction for the beam incidence is parallel to the axis of a dodecagonal atomic column.  On the other hand, we can see a lot of fine, dark bands due to structural disorders in Area II, as indicated by the arrows in Fig. 2(b).  A prominent feature of the pattern shown in the inset in Fig. 2(b) is that reflections with lower scattering angles around the origin 000 have relatively weak intensities, while twelve strong reflections are present around the scattering angle of $\sin\theta/\lambda = 1.22 \; nm^{-1}$, as indicated by the twelve short arrows.  The point to note here is that this scattering angle corresponds to the diameter of a dodecagonal atomic column. This implies that an array of dodecagonal atomic columns present in the H structure has a short-range nature in Area II.

To understand the features of dodecagonal column arrangements in Areas I and II, high-resolution electron micrographs were taken from the two areas. Figure 3 shows a $[0001]_H$ high-resolution electron micrograph taken from Area I, together with the corresponding electron diffraction pattern in the inset in Fig. 3(a), and a calculated micrograph of the H structure in Fig. 3(b). A projection of the atomic positons in the H structure along the $[0001]_H$ direction is also depicted in Fig. 3(b). The calculated micrograph was obtained under the conditions of a defocus value of about -20 nm and a sample thickness of about 27 nm, and the crystallographic data of the H structure reported by Iga *et al.* were used for the calculation [18]. In the experimental micrograph for Area I, we can clearly see a regular array of bright dots with hexagonal symmetry. To investigate the origin of the bright dots, the experimental micrograph was compared with both the calculated micrograph and the projected atomic positions in the H structure along the $[0001]_H$ direction. It was found that the calculated micrograph well reproduces the experimental one, and that each bright dot corresponds to the location of the axis of a dodecagonal atomic column. In other words, the array of bright dots observed in the experimental micrograph basically reflects that of dodecagonal atomic columns. As reported in a previous study, the experimental micrograph also exhibits a periodic array of dodecagonal structural units indicated by the white open circles, in addition to the dodecagonal atomic columns [18]. Note that each dodecagonal structural unit consists of 19 dodecagonal atomic columns with a hexagonal configuration. A notable feature of the array of dodecagonal structural units is that two neighboring units share two dodecagonal atomic columns, as shown in the purple region in Fig. 3(b). This indicates that information on the arrays of both dodecagonal structural units and dodecagonal atomic columns can be obtained from such an array of bright dots observed in experimental micrographs. With the help of the information on the array of bright dots, we tried to identify the features of the arrays of the dodecagonal structural units and dodecagonal atomic columns present in Area II.

Figure 4 shows two $[0001]_H$ high-resolution electron micrographs taken from two different regions in Area II. These two micrographs were obtained under the same conditions as that for Area I, and each white open circle represents a dodecagonal structural unit. We can see various arrays of bright dots indicating the axis positions of dodecagonal atomic columns, although they are not detected in some regions indicated by the white arrows. The point to note here is that an array of bright dots for the σ structure can also be found in the region marked by the letter σ. The green lines in the σ region indicate an array of dodecagonal atomic columns in the σ structure, as was reported from our previous study [22]. The dodecagonal structural units indicated by the white open circles in Fig. 4(a) have two typical types of penetrated structural units, referred to as the PUI and PUII units, as structural disorders, which consist of two and seven dodecagonal structural units. Schematic diagrams of the PUI and PUII units are depicted in the inset in Fig. 4(a). The diagrams of these two penetrated structural units show that two neighboring units share six dodecagonal atomic columns, as marked by the transparent red regions. Another interesting feature is that three neighboring units present in the PUII unit have three common columns, as indicated as the transparent blue regions. In addition to the penetrated structural units, we can find a square configuration of four dodecagonal structural units, as indicated by the brown arrows, as well as the hexagonal configuration of seven dodecagonal units in an H region on the lower side of Fig. 4(a). In Fig. 4(b), there are two H regions with different orientations, which are referred to as the H1 and H2 orientations, in addition to σ regions with a single orientation Σ. This is apparently a direct experimental indication that there are two kinds of orientation relations between the σ and H structures. Figure 4(c) shows two electron diffraction patterns with the H1 and H2 orientations for the H structure, together with a $[001]_\sigma$ pattern of the σ structure with the Σ orientation. These patterns were taken from two different regions of a relatively large size, and the pattern for the H2 orientation was obtained merely by the counterclockwise rotation of the pattern for the H1 orientation by 30° about the $[0001]_H$ direction. As indicated by the small open circles in the patterns, the location of the $140_\sigma$ reflection for the σ structure is found to be coincident with that of the $\bar{4}8\bar{4}0_H$ reflection for the H1 orientation and that of the $07\bar{7}0_H$ reflection for the H2 orientation. This implies that there are two kinds of orientation relations of $(140)_\sigma // (\bar{4}8\bar{4}0)_H$ and $[001]_\sigma // [0001]_H$, and $(140)_\sigma // (07\bar{7}0)_H$ and $[001]_\sigma // [0001]_H$ between the σ and approximant H structures.

B. Mn-17at.%Si-9at.%V alloy samples

Based on the above-mentioned experimental data, the features of the (σ → H) reaction were examined by using Mn-17at.%Si-9at.%V samples. Figure 5 shows three x-ray powder diffraction profiles in the angular range of $35° \leq 2\theta \leq 55°$, which were measured from three alloy samples subjected to different thermal treatments. The sample for the profile in Fig. 5(a) was kept at 1273 K for 24 h as the solution treatment to obtain the single σ phase as the starting state of the reaction. To induce the (σ → H) reaction, the samples for the profiles in Figs. 5(b) and 5(c) were, respectively, annealed at 1027 K for 8 h and 100 h as the thermal treatment following the solution treatment. We first look at the profile in Fig. 5(a) for only the solution treatment.  The profile exhibits sharp reflections with clear splitting of the reflections due to Cu$K\alpha_1$ and Cu$K\alpha_2$ radiation.  A simple analysis of the profile indicated that the reflections were entirely consistent with those due to the σ structure, except for the ones marked by the black arrows.  Although unknown black-arrow reflections are present in the profile, the sample basically consisted of σ structure regions as the starting state of the reaction.  For the sample kept at 1073 K for 8 h, its profile in Fig. 5(b) is basically identical to that of the starting σ state in Fig. 5(a).  The differences between the two profiles are the absence of the unknown reflections and the appearance of new reflections with weak intensities, as indicated by the blue arrows.  For the sample annealed at 1073 K for 100 h, the new blue arrow reflections became much stronger, as shown in Fig. 5(c).  Because the profile in Fig. 5(c) is identical to that of the hexagonal H structure in Fig. 1, the sample annealed at 1073 K for 100 h can thus be identified as the approximant H state representing the final state in the reaction.  Based on these results, the (σ → H) reaction was confirmed to actually occur in the samples with the Mn-17at.%Si-9at.%V composition.

The measured x-ray powder diffraction profiles indicated that the (σ → H) reaction occurred by keeping the sample with the metastable σ state at 1073 K.  The detailed features of the (σ → H) reaction were then examined by taking high-resolution electron micrographs of samples consisting of the (σ + H) coexistence state.  Figures 6(a), 6(b), and 6(c) show high-resolution electron micrographs with the [001]$_\sigma$ electron beam incidence, which were taken from three different areas of the same sample annealed at 1073 K for 8 h, together with a micrograph of a sample annealed for 100 h in Fig. 6(d). The three micrographs for 8-hour annealing are arranged in the order of the increasing volume of the H regions, and the four regions shown here are referred to as Regions I, II, III, and IV.  The increase in volume presumably reflects the progress of the (σ → H) structural change.  These four micrographs were taken under the same experimental conditions as those shown in Fig. 3(a) and Fig. 4.  Dodecagonal structural units in the micrographs are again indicated by the white open circles.  Among these four micrographs, there is no H region in Region I, where a thin band consisting of penetrated structural units formed along one of the <110>$_\sigma$ directions in the σ matrix.  This is an experimental indication that the (σ → H) structural change starts with the nucleation of a penetrated structural unit consisting of dodecagonal structural units in the σ matrix.  In Region II, we can clearly see a rectangular arrangement of PUI units in a large region, referred to as the RAPU region.  The notable features of Region II are that σ regions still remain as fine bands in the interior of the RAPU region, as indicated by the green arrows, and that an H region is present in contact with the RAPU region, as shown by the red arrow. These features of Region II suggest that RAPU regions act as a nucleation site for the formation of H regions.

Electron diffraction patterns were taken at various beam incidences for RAPU regions. An analysis of the experimentally obtained patterns indicated that the RAPU state has orthorhombic Pmmm symmetry, where the unit cell of the RAPU state is marked by the black square in the Fig. 6(b). To understand the crystallographic relation between the σ and RAPU states, the $[001]_R$ pattern shown in the inset in Fig. 6(b) was taken from an area including the σ and RAPU regions. The overlapped pattern in the inset clearly exhibits a simple orientation relationship of $(110)_\sigma // (200)_R$ and $[001]_\sigma // [001]_R$ between the σ and RAPU states. In addition, the appearance of the RAPU state from the σ one was accompanied by a lattice contraction with a magnitude of about (10/11 ~ 0.91%) along the $[1\bar{1}0]_\sigma$ direction, as indicated by the arrows in the pattern. As a result, the lattice parameters of the orthorhombic RAPU state are given by $a^R \sim 2d_{110}^\sigma$, $b^R \sim 3d_{110}^\sigma \times (10/11)$, and $c^R \sim d_{001}^\sigma$. In other words, no lattice distortion was basically found along the $[110]_\sigma$ and $[001]_\sigma$ directions.

In Region III, two relatively large H regions are present in the upper left and lower right sides of the micrograph, in addition to large RAPU and small σ regions. The striking feature of the RAPU regions is that they exhibit a domain structure, and arrays of PUI units in two neighboring RAPU domains are characterized by an out-of-phase shift along one of the <110>$_\sigma$ directions, as indicated by the brown lines. In addition, the H region in the lower right side serves as an out-of-phase region between two neighboring RAPU regions. As seen in the micrograph for Region IV, the approximant H state is also found to be an assembly of out-of-phase domains with a phase shift of $2\pi/3$ along one of the <01$\bar{1}$0>$_H$ directions. It seems that the out-of-phase domain structure in the approximant H state follows that in the RAPU regions. One of the interesting features in the domain structure of the H state is a one-dimensional array of dodecagonal structural units with a square configuration, which may act as an out-of-phase boundary in the H structure, as indicated by the small arrows.

4. Discussion

In the present study, two kinds of Mn-Si-V alloy samples with the compositions of Mn-20at.%Si-10at.%V and Mn-17at.%Si-9at.%V were used to investigate the crystallographic features of both the approximant H phase and the (σ → H) reaction. It was found that there were two kinds of areas (Areas I and II) in Mn-20at.%Si-10at.%V samples. Concretely, uniform H structure regions were present in Area I, while the H structure in Area II displayed structural disorders. The major disorders were characterized by the presence of the penetrated structural units; that is, the PUI and PUII units. In addition, the square configuration of four dodecagonal structural units and local σ regions were also detected in Area II, together with the hexagonal configuration of seven dodecagonal units in approximant H regions. With the help of the coexistence of σ and H regions, two kinds of orientation relations of $(140)_\sigma$ // $(\bar{4}8\bar{4}0)_H$ and $[001]_\sigma$ // $[0001]_H$, and $(140)_\sigma$ // $(07\bar{7}0)_H$ and $[001]_\sigma$ // $[0001]_H$ were further established between the σ and approximant H structures. The (σ → H) reaction, examined by using Mn-17at.%Si-9at.%V samples, started with the appearance of PUI units, and the spatial growth of PUI units resulted in RAPU regions as a metastable state. Approximant H regions were then nucleated from the σ matrix in contact with RAPU regions. In other words, RAPU regions were likely to act as a nucleation site for the formation of H regions. Based on these experimental data, we will here discuss the crystallographic features of the formation of the approximant H structure from the σ structure via the metastable RAPU state. In addition, atomic positions in both the approximant H structure and the dodecagonal quasicrystal will also be discussed based on the simple plane-wave model with twelve plane waves. It will be noted that a similar plane-wave model was previously used in our analysis of the decagonal quasicrystal in the Al-Co-Cu alloy system [23].

The detailed features of the (σ → H) reaction via the metastable RAPU state are discussed here on the basis of our experimentally obtained data. To understand the role of the RAPU state as a metastable state, the discussion starts with an analysis of the direct (σ → H) structural change with the help of the determined orientation relations. Previously reported atomic positions for the σ and H structures were first overlapped on the basis of the determined relation of $(140)_\sigma$ // $(\bar{4}8\bar{4}0)_H$ and $[001]_\sigma$ // $[0001]_H$. Even when lattice distortions were taken into account, a one-to-one correspondence between them could not be obtained. This implies that the appearance of the RAPU state apparently helps the formation of the H structure. The important points of the RAPU state are that the orientation relation between the σ and RAPU states is given by $(110)_\sigma$ // $(200)_R$ and $[001]_\sigma$ // $[001]_R$, and that no lattice distortion is found along the $[110]_\sigma$ and $[001]_\sigma$ directions. The unit cell is estimated to be $a^R \sim 2d^\sigma_{110} \sim 1.25\ nm$, $b^R \sim 3d^\sigma_{110} \times (10/11) \sim 1.702\ nm$, and $c^R \sim d^\sigma_{001} \sim 0.463\ nm$ on the basis of the crystallographic data reported by Dickins *et al.* for the tetragonal P4$_2$/mnm σ structure [21]. Note that the factor (10/11) for the lattice parameter $b^R$ is due to the lattice contraction along the $[1\bar{1}0]_\sigma$ direction. Figure 7 shows both a projection of atomic positions in the σ and RAPU states along the $[001]_\sigma$ direction and a three-dimensional diagram for a part of the σ structure. In the diagram, atoms in the σ and RAPU states are, respectively, represented by the blue and red circles. As seen in the diagram, a one-to-one correspondence between atomic positions in the σ and RAPU states can be established in a banded region with a width of about 3 nm, as indicated by the double arrows. In the banded region, atomic positions in the RAPU state are obtained from those in the σ structure by simple atomic shifts mainly along the $[001]_\sigma$ direction, as indicated by the thick black arrow in the three-dimensional diagram. This implies that banded RAPU regions can be formed by simple shifts between two neighboring layers, basically with no lattice distortion. With the appearance of the RAPU state, dodecagonal structural units characterizing the approximant H structure are formed by simple atomic shifts basically along the column axis.

In the (σ → H) reaction, H regions were nucleated from the σ matrix in contact with RAPU regions. This suggests that RAPU regions are likely to act as a nucleation site for the formation of H regions. We then checked the lattice parameters in the RAPU state and the H structure. Based on the orientation relations both between the σ and H states and between the σ and RAPU states, the relation of $(020)_R$ // $(\bar{2}110)_H$ and $[001]_R$ // $[0001]_H$ is found to be present between the RAPU state and the H structure. According to the crystallographic data reported by Iga *et al.* for the hexagonal P4/mmm H structure, it is also found that the lattice parameters of $a^H = 1.7058\ nm$ and $c^H = 0.4640\ nm$ are almost equal to $b^R \sim 1.702$ nm and $c^R \sim 0.463$ nm in the RAPU state, respectively. This implies that H and RAPU regions can coherently contact each other in the $(100)_R$ (= $(01\bar{1}0)_H$) plane. This would explain why RAPU regions can act as a nucleation site for the formation of H regions.

A conversion of the RAPU state into the H state is needed for the growth of nucleated H regions. We then checked atomic shifts in the change from the RAPU state to the H structure. Figure 8 shows the column axis projections of atoms in both the RAPU and H states together with a three-dimensional diagram for a part of the RAPU state. The red and yellow circles in the diagram represent atoms in the RAPU and H states. It is seen in the diagram that one-to-one correspondences are established between atomic positons in a banded region indicated by the double arrows, where the (RAPU →

H) change can occur by simple atomic shifts indicated by the black arrows in the three-dimensional diagram, basically along the column axis direction. The (σ → H) change via the RAPU state can thus be understood as an effort by alloys to avoid lattice distortions, and it can occur only by simple atomic shifts basically along the column axis direction. It will be noted again that, instead of the appearance of the approximant H state, the metastable RAPU state results in the formation of dodecagonal structural units, which are directly associated with that of the dodecagonal quasicrystal.

In our previous study on the decagonal quasicrystal, we proposed a stationary-wave model using ten plane waves on the basis of the Hume-Rothery mechanism [23]. The model was found to well predict the features of the decagonal quasicrystal in the Al-Co-Ni alloy system such as electron diffraction patterns and atomic positions. To verify the generality of our model, in this study, a similar model was applied to predict atomic positions of both the approximant H structure and the dodecagonal quasicrystal. Here, the following point concerning the dodecagonal quasicrystal should be noted. Dodecagonal quasicrystals have so far been found in macromolecular polymers and colloids, in addition to alloys such as Mn-Si-V alloys [1–11]. This implies that the physical origin of the appearance of the dodecagonal quasicrystal is not restricted to the Hume-Rothery mechanism. Based on only the fact that the features of the decagonal quasicrystal were reproduced by using the ten plane waves, we simply predicted the atomic positions by superposing twelve plane waves with a common crystallographic component, which produced the pseudo-twelvefold symmetry in the H structure. In other words, we are not concerned about the physical origin for the appearance of the approximant H phase and the dodecagonal quasicrystal. Figure 9 shows two electron diffraction patterns of the approximant H structure, which were taken from a Mn-17at.%Si-9at.%V sample annealed at 1073 K for 100 h. The electron beam incidences of the patterns in Figs. 9(a) and 9(b) are, respectively, parallel to the $[1\bar{2}10]_H$ and $[0001]_H$ directions in terms of the H structure notation. One of the twelve wave vectors adopted in the present study is shown in the pattern with the $[1\bar{2}10]_H$ incidence in Fig. 9(a). This vector is assumed to be the reciprocal lattice vector for the $70\bar{7}1_H$ reflection, which is indicated by the long blue arrow. Its feature is that the vector has both the two-dimensional pseudo-quasi-periodicity in the $(0001)_H$ plane and the one-dimensional crystallographic periodicity along the pseudo-twelvefold $[0001]_H$ axis at the same time. When the vector is referred to here as $\boldsymbol{k}_1^H$, the crystallographic and pseudo-quasi-periodic components are, respectively, given by $\boldsymbol{k}_1^{HC} = \boldsymbol{k}_1^H \sin\theta$ and $\boldsymbol{k}_1^{HP} = \boldsymbol{k}_1^H \cos\theta$ with $\theta \sim 25°$. The crystallographic periodicity for $\boldsymbol{k}_1^H \sin\theta$ is apparently equal to the lattice parameter of $c^H = 0.4640\ nm$ in the H structure. The pseudo-quasi-periodic components of the adopted twelve waves, $k_j^H$ with integer $j$ from 1 to 12, are indicated by the twelve green arrows in the pattern with the $[0001]_H$ incidence in Fig. 9(b). As shown in the pattern, the angle between two neighboring vectors was found to be 30°, and the magnitudes of the twelve components were determined to be 29.78 nm$^{-1}$ for $j = 1$, 3, 5, 7, 9, and 11, and 29.47 nm$^{-1}$ for $j = 2, 4, 6, 8, 10$, and 12. Because there is the slight difference in their magnitudes, the H structure has hexagonal symmetry. We can thus obtain twelvefold symmetry in the dodecagonal quasicrystal by assuming the same magnitudes of these twelve waves.

In our plane-wave model, the atomic positions in both the approximant H structure and the dodecagonal quasicrystal were predicted simply by superposing the twelve plane waves with the wave vectors $k_j^H$. We first write the $j$-th plane wave as $\varphi_j(\mathbf{r}) = \varphi_{0j} \exp\{i(k_j^H) \cdot \mathbf{r}\}$ with a complex amplitude $\varphi_{0j} = |\varphi_{0j}| \exp(i\theta_j)$. As a result, the total real wave $P(\mathbf{r})$ containing the twelve plane waves is given by the following simple form,

$$P(\mathbf{r}) = \sum_{j=1}^{12} A_j \cos(k_j^H \cdot \mathbf{r} + \theta_j) \quad (1),$$

where $A_j = 2|\varphi_{0j}|$. Here, the following point should be noted. Both the approximant H structure and the dodecagonal quasicrystal consist of four layers, whose positions are located at $z = 0, 1/4, 1/2$, and $3/4$ by using the z component of the coordinates. We then focus on the atomic arrangements in these four layers. Next, the wave vectors $k_j^H$ was decomposed into two components; that is, the crystallographic and pseudo-quasi-periodic components, $k_j^H = k_j^{HP} + k_j^{HC}$, and the term of $k_1^{HC} \cdot \mathbf{r} = 2\pi z$ was calculated for $z = 0$, $1/4, 1/2$, and $3/4$. As a result, Eq. (1) may be written as

$$P(\mathbf{r}) = \sum_{j=1}^{12} A_j \cos(k_j^{HP} \cdot \mathbf{r} + \Delta_j) \quad (2),$$

where the phase $\Delta_j$ is given by $\theta_j$ for $z = 0$, $\theta_j + \pi/2$ for $z = 1/4$, $\theta_j + \pi$ for $z = 1/2$, and $\theta_j + 3\pi/2$ for $z = 3/4$. The phase $\Delta_j$ apparently includes the contribution of the crystallographic component along the $[0001]_H$ direction. In addition, the assumption $A_j = 1$ was made for simplicity in the present calculation. Thus, our plane-wave model has only one adjustable parameter $\Delta_j$ for obtaining atomic positions in both the approximant H structure and the dodecagonal quasicrystal.

To determine values of $\Delta_j$ for the four layers, we first calculated the distribution of the total real wave $P(\mathbf{r})$ for each layer by using Eq. (2), and the calculated distribution was compared with atomic positions in the approximant H structure. The crystallographic data reported by Iga *et al.* were used for determining the atomic positions in the H structure [18].   The calculated distributions for the four layers at z = 0, 1/4, 1/2, and 3/4 are shown in Fig. 10 together with both the atomic positions in the H structure on the right-hand side of each distribution and the values of $\Delta_j$ determined for each wave in the inset.   The point to note here is that the atomic positions in the layers at z = 0 and 1/2 are identical, as shown in the left-side distribution in the figure. In the distributions, the magnitude of the $P(\boldsymbol{r})$ value decreases in order of the orange, faint orange, faint blue, and blue colors.   A comparison between the calculated distributions and atomic positions indicated by the small yellow circles reveals that the distributions reproduce the atomic positions in the H structure by putting atoms at the positions with $P(\boldsymbol{r})$ values above the threshold values of 9.32 for z = 0 and 1/2, 6.00 for z = 1/4, and 6.00 for z = 3/4. In other words, atomic positions in the approximant H structure can well be explained by our simple plane-wave model.   Based on this, we tried to predict atomic positions in the dodecagonal quasicrystal with our model using the same $\Delta_j$ and threshold values as those determined in the approximant H structure.

　　In the calculation, the dodecagonal quasicrystal can be obtained from the H structure by assuming the same magnitudes of $\boldsymbol{k}_j^{HP}$ for the twelve waves to get twelvefold symmetry.   The value of 29.78 nm$^{-1}$ for $\boldsymbol{k}_j^{HP}$ with $j$ = 1, 3, 5, 7, 9, and 11 in the H structure was actually used for the magnitude of all $\boldsymbol{k}_j^{DQ}$ in the dodecagonal quasicrystal.   The calculated distributions of the $P(\boldsymbol{r})$ value for the layers at z = 0 and 1/2, 1/4, and 3/4 are shown in Fig 11 together with the predicted atomic positions in the dodecagonal quasicrystal, which are indicated by the green dots.   In the figure, we can actually see an arrangement of atoms with twelvefold symmetry.   From the calculated arrangement of atoms shown in Fig. 11, however, it is hard to understand the crystallographic features of the dodecagonal quasicrystal.   A tiling pattern of the dodecagonal quasicrystal was thus calculated by using the atomic positions in the layer at z = 0, which are shown in Fig. 11.   Because each atomic position in the z = 0 layer corresponds to that of a column axis in a dodecagonal atomic column, a tiling pattern was concretely constructed by connecting nearest-neighbor atoms at distances from 0.452 nm to 0.456 nm.   A tiling pattern of the dodecagonal quasicrystal obtained in this way is depicted in Fig. 12 together with four types of tiles found in the pattern.   The point to note here is that six of the twelve lattice points around the center of the diagram were omitted to avoid gathering lattice points with shorter distances. The resultant hexagon consisting of six triangles is shown at the center of Fig. 12, as indicated by the arrow. As seen in the diagram, the pattern is composed of square-, triangle-, rhombus-, and distorted hexagonal-shaped tiles, as indicated by the green, blue, red, and yellow colors, respectively. Among these four types of tiles, the distorted hexagonal-shaped tile is found to be an assembly of two triangles, one square, and one rhombus, as shown in the lowest inset. This is an indication that the tiling pattern of the dodecagonal quasicrystal basically consists of triangle-, square-, and rhombus-shaped tiles, as already reported in previous studies. Because the tiling pattern of the H structure can be expressed by triangle- and square-shaped tiles, the presence of a rhombus-shaped tile is one of the characteristic features of the tiling pattern of the dodecagonal quasicrystal. Another interesting feature is that when an edge in a rhombus-shaped tile is assumed to be a unity, the length of a diagonal line $\Omega$ is given by

$$\Omega = 2\cos 15° = \frac{\sqrt{3}+1}{\sqrt{2}} \sim 1.932.$$

In addition, we can get $\Omega^2 = 2 + \sqrt{3}$ and $\Omega^4 = 7 + 4\sqrt{3}$. This suggests that the value of $\Omega$ is a mathematical constant just like the golden number $\tau$. We believe that the constant $\Omega$ characterizes self-similarity, which is present in the atomic arrangement in the dodecagonal quasicrystal.

5. Conclusions

In this study, the crystallographic features of both the approximant H phase for the dodecagonal quasicrystal and the (σ → H) reaction in the Mn-Si-V alloy system were investigated mainly by transmission electron microscopy. The results confirmed that the crystal structure of the H phase is characterized by a regular array of dodecagonal structural units consisting of 19 dodecagonal atomic columns, where the dodecagonal atomic column is formed by connecting CN14 polyhedra along a one-dimensional direction. Typical examples of structural disorders in the H structure were identified as penetrated structural units and a square configuration of four dodecagonal structural units. In addition, the (σ → H) reaction started with the appearance of regions characterized by a rectangular arrangement of penetrated structural units (RAPU), and H regions were then nucleated from the σ matrix in contact with RAPU regions. In other words, dodecagonal structural units were formed, together with the appearance of the RAPU state, prior to that of the H state. The subsequent growth of H regions at the expense of RAPU regions finally resulted in the approximant H state. Furthermore, atomic positions in the dodecagonal quasicrystal were also predicted with a simple plane-wave model consisting of twelve plane waves.

Acknowledgement

This study was partially supported by the Japan Society for the Promotion of Science (JSPS) in the form of a Grant-in-Aid for Scientific Research (C), 15K06426, 2015.

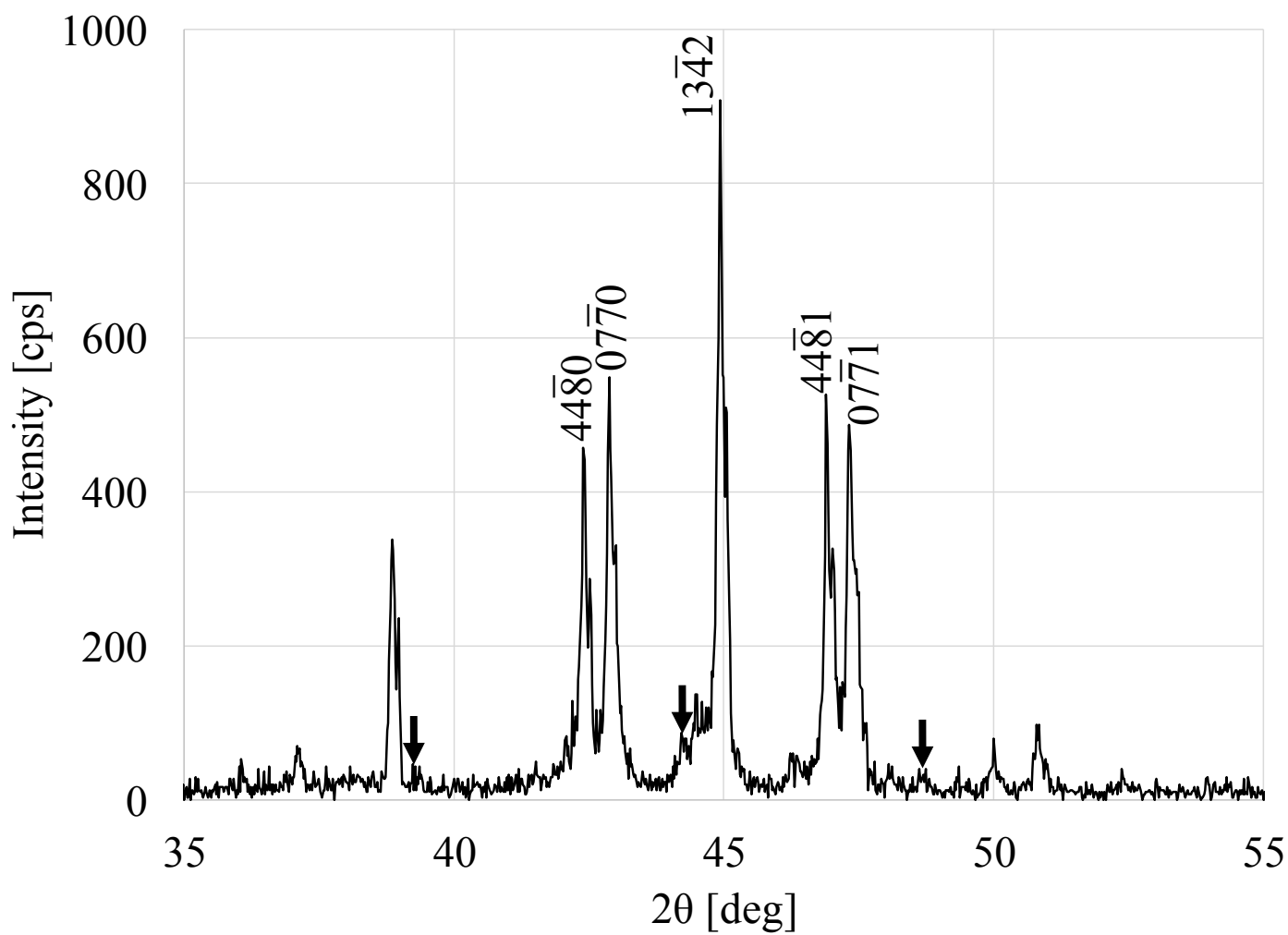

Fig. 1. Measured x-ray powder diffraction profile of a Mn-20at.%Si-10at.%V alloy sample at room temperature in the angular range of $35° \leq 2\theta \leq 55°$. The profile was confirmed to be basically consistent with the approximant H structure with hexagonal P6/mmm symmetry. In addition to the H reflections, reflections with weak intensities were also detected in the profile, as indicated by the black arrows.

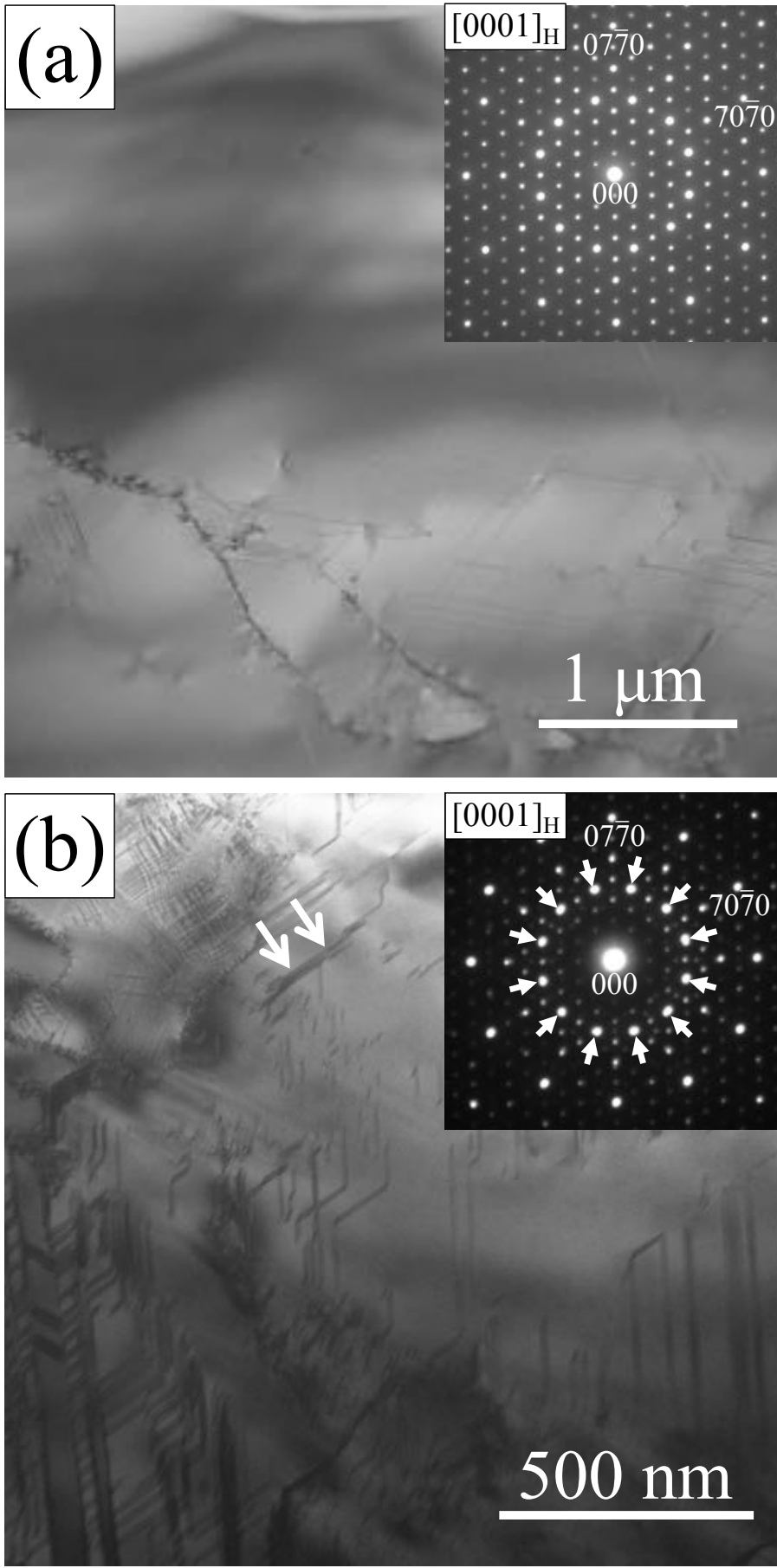

Fig. 2. Two bright-field images taken from two different areas in the Mn-20at.%Si-10at.%V alloy sample at room temperature, together with corresponding electron diffraction patterns in the insets. The electron beam incidence for these two images is parallel to the $[0001]_H$ direction in the hexagonal notation, and these two areas are here referred to as Areas I and II. From the pattern in (a) for Area I, its state exhibiting a relatively uniform contrast in the image is confirmed to be the hexagonal P6/mmm H state. On the other hand, a lot of fine dark bands due to structural disorders are seen in (b) for Area II, as indicated by the arrows. Because twelve reflections around $\sin\theta/\lambda = 1.22\ nm^{-1}$ have stronger intensities in the pattern for Area II, as indicated by the short arrows, it is inferred that an array of dodecagonal atomic columns in the H structure should be of a short-range nature in Area II.

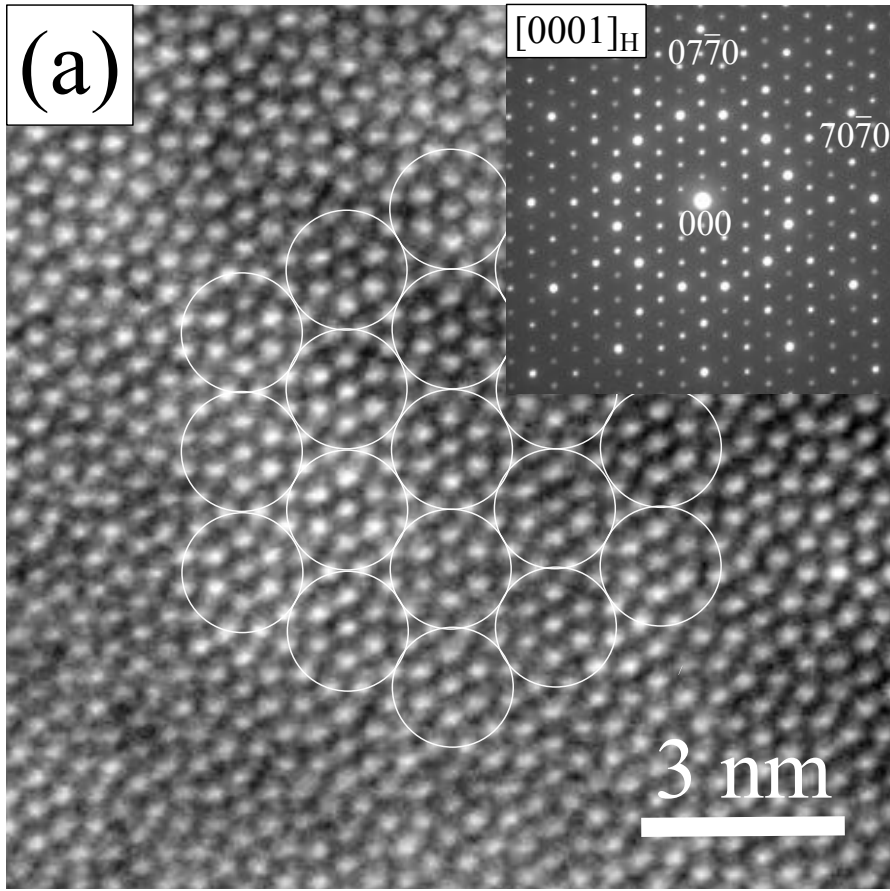
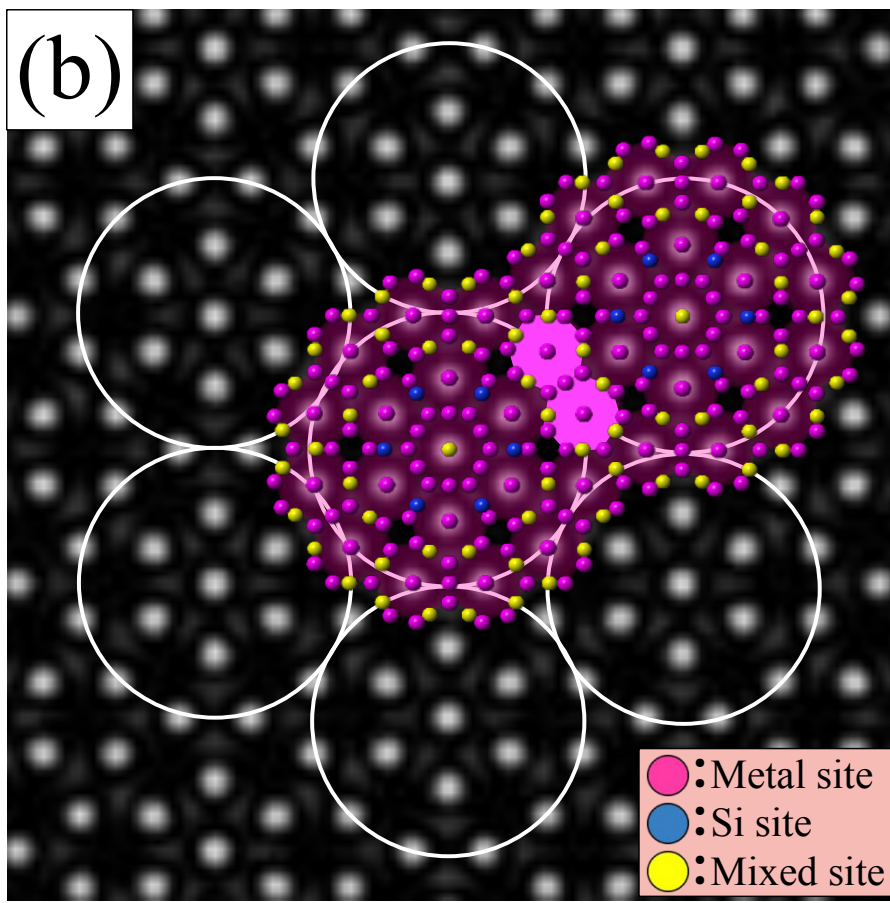

Fig. 3. High-resolution electron micrographs taken from Area I in the Mn-20at.%Si-10at.%V alloy sample and a calculated micrograph of the hexagonal P6/mmm H structure, together with a corresponding $[0001]_H$ electron diffraction pattern. The calculated micrograph in (b) was obtained under the conditions of a defocus value of about -20 nm and a sample thickness of about 27 nm, and the crystallographic data of the H structure reported by Iga *et al.* were also used in the calculation. The calculated micrograph in (b) well reproduces the experimental one in (a), and each bright dot corresponds to the location of the axis of a dodecagonal atomic column. In addition, the experimental micrograph exhibits a periodic array of dodecagonal structural units consisting of 19 dodecagonal atomic columns, as indicated by the white open circles.

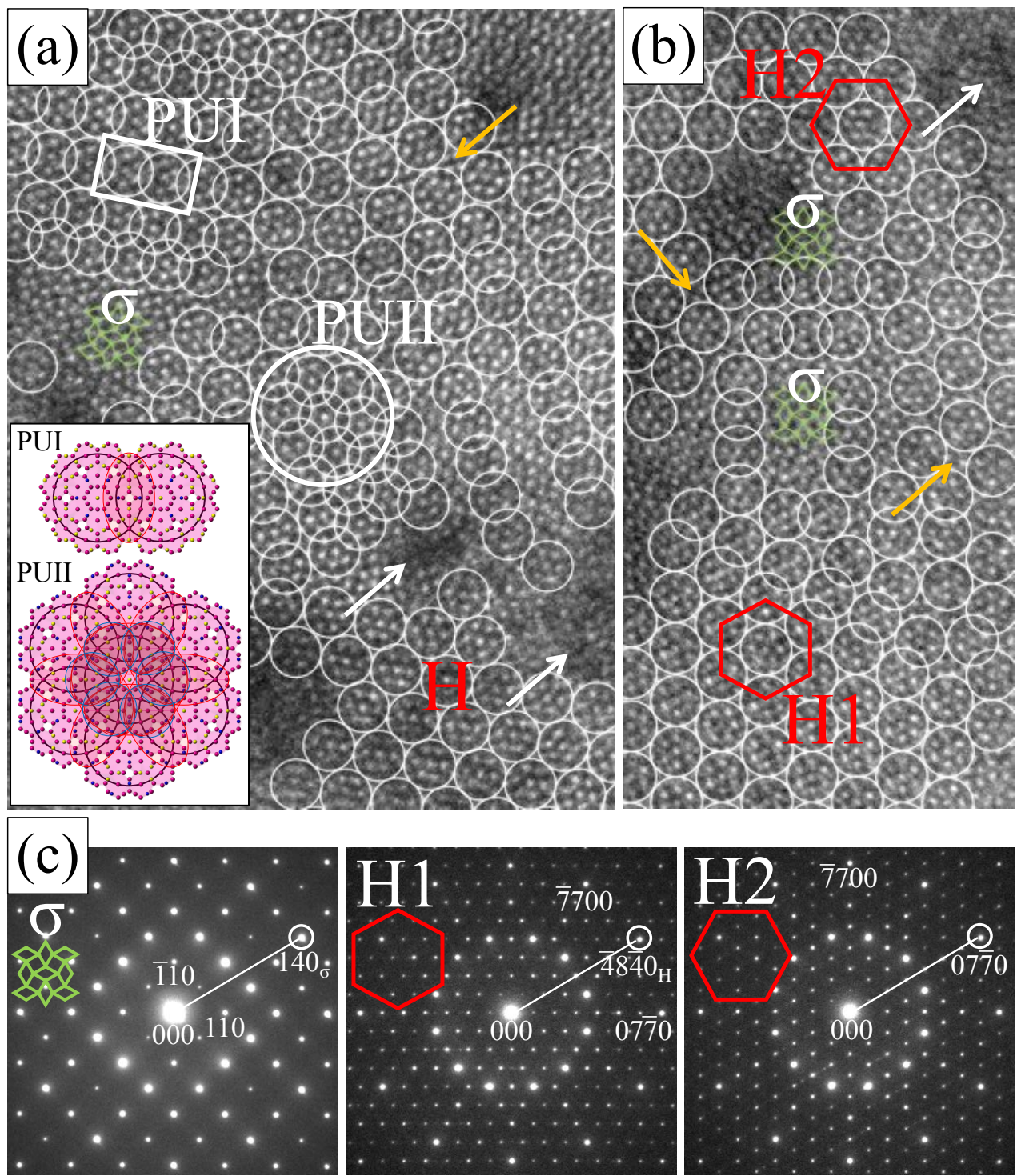

Fig. 4. Two $[0001]_H$ high-resolution electron micrographs taken from two different regions in Area II of the Mn-20at.%Si-10at.%V alloy sample, together with electron diffraction patterns indicating orientation relations between the σ and approximant H structures. The experimental condition for the two micrographs was the same as that for Area I in Fig. 3(a), and each white open circle represents a dodecagonal structural unit. Various arrays of bright dots indicating axis positions of dodecagonal atomic columns are seen in the micrographs, and an array of bright dots for the σ structure is also found in a region marked by letter σ. For the dodecagonal structural units, penetrated structural units referred to as the PUI and PUII units are present as structural disorders in (a). In addition, there are two H regions with the H1 and H2 orientations in (b), in addition to σ regions with a single orientation. Based on a $[001]_\sigma$ electron diffraction pattern for the σ structure with a single orientation and two $[0001]_H$ patterns for the H1 and H2 orientations in (c), two kinds of the orientation relations of $(140)_\sigma // (\bar{4}8\bar{4}0)_H$ and $[001]_\sigma // [0001]_H$, and $(140)_\sigma // (07\bar{7}0)_H$ and $[001]_\sigma // [0001]_H$ are established between the σ and H structures.

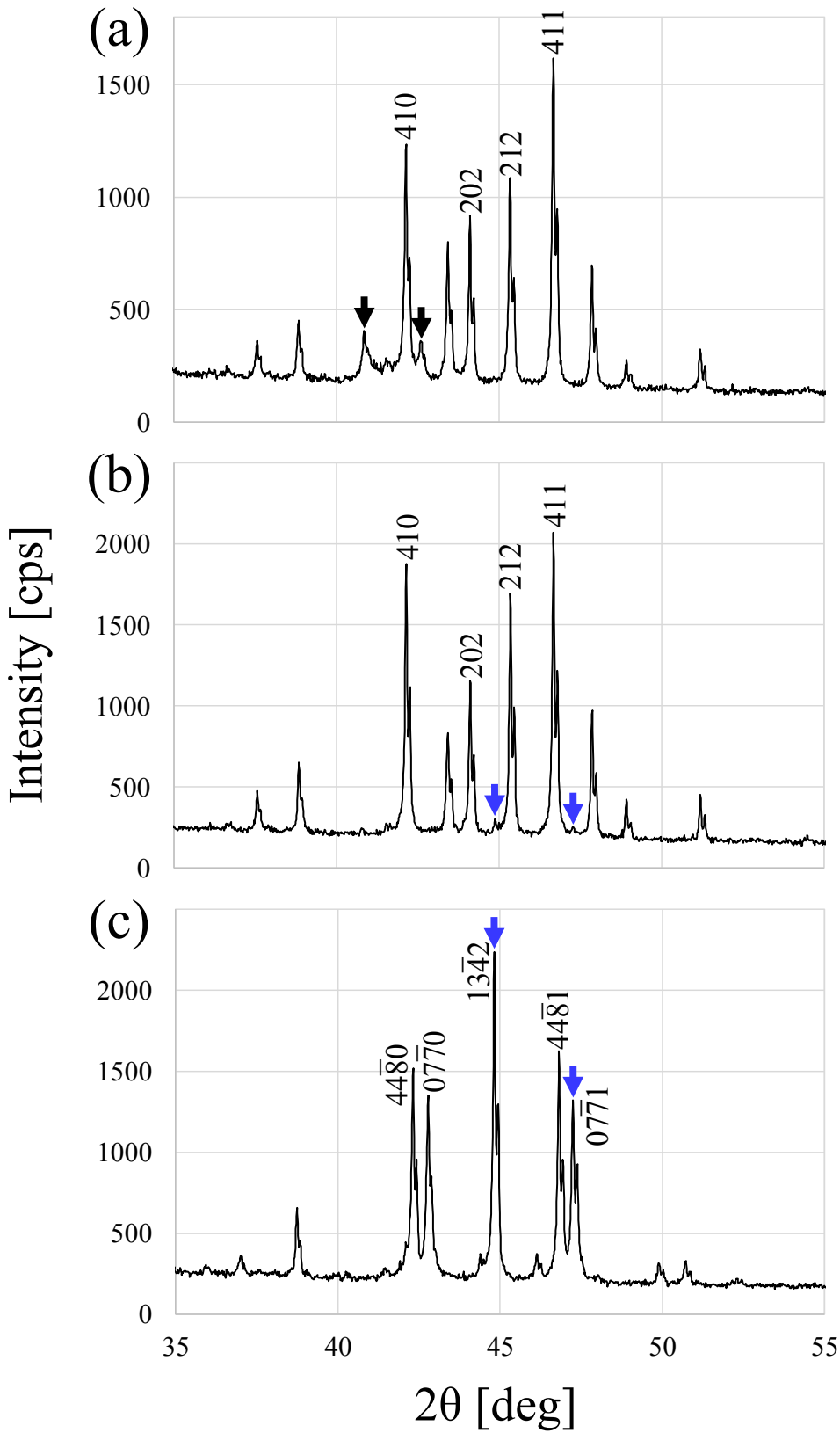

Fig. 5. Three x-ray powder diffraction profiles at room temperature in the angular range of 35° ≤ 2θ ≤ 55°, measured from three Mn-17at.%Si-9at.%V alloy samples with different thermal treatment conditions. The sample for (a) was kept at 1273 K for 24 h to obtain the single σ state as the starting state. The samples for (b) and (c) were, respectively, annealed at 1027 K for 8 h and 100 h, respectively, to induce the (σ → H) reaction. The profile in (a) indicates that, in spite of the presence of unknown reflections marked by the black arrows, the reflections are basically consistent with those due to the σ structure. From the profiles in (b) and (c), reflections indicated by the blue arrows appear and become much stronger with an increase in annealing time. Because there is no σ reflection in (c), the profile is found to be identical to that of the approximant H structure.

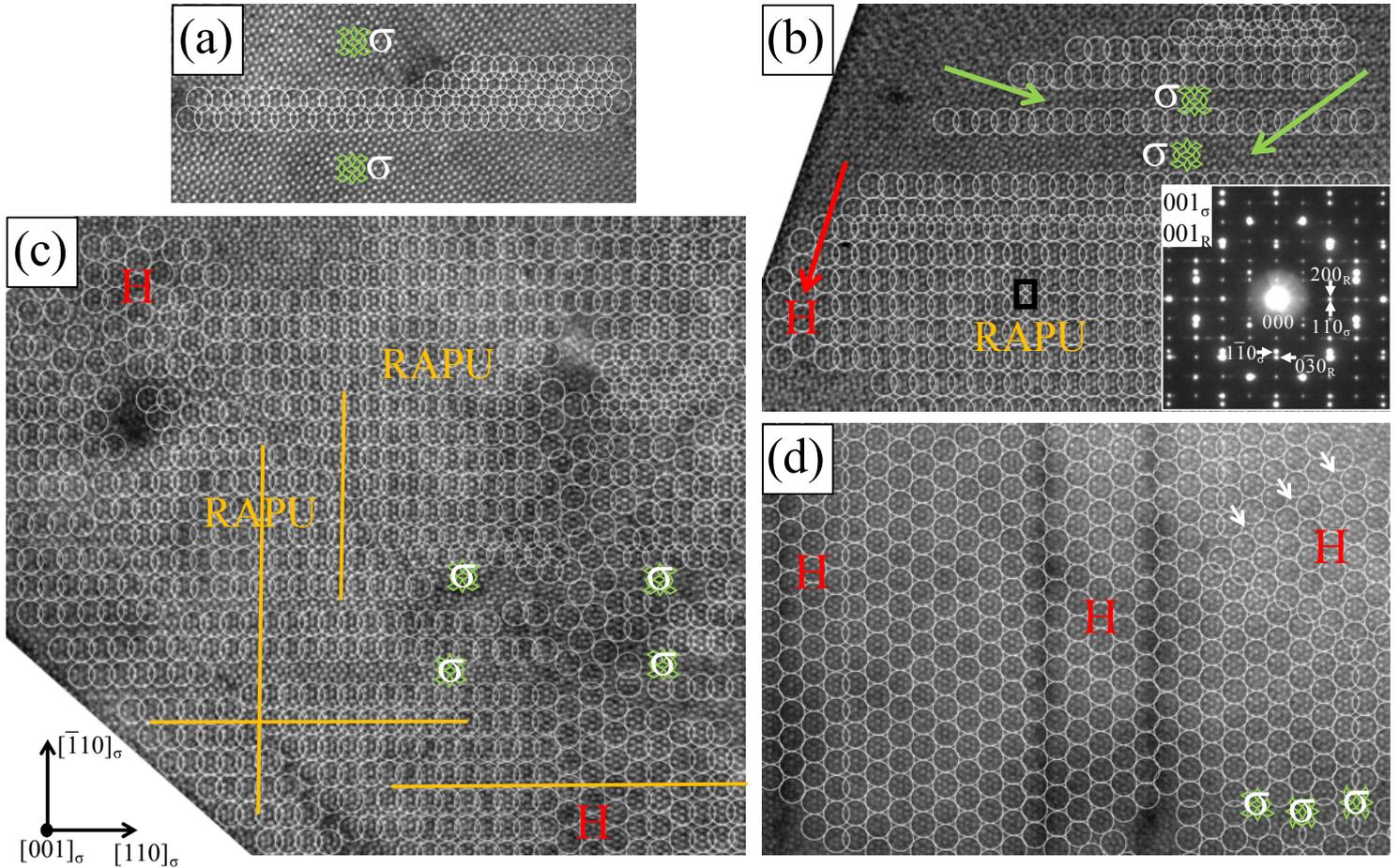

Fig. 6. $[001]_\sigma$ high-resolution electron micrographs of four different areas in the Mn-17at.%Si-9at.%V alloy samples at room temperature. Concretely, the micrographs in (a), (b), and (c) for the (σ + H) coexistence state were obtained from the sample annealed at 1073 K for 8 h, while the H structure sample annealed at 1073 K for 100 h exhibited the micrograph in (d). The three micrographs for the 8-hour annealed sample are arranged in increasing volume of H regions, and the four regions for (a), (b), (c), and (d) are referred to as Regions I, II, III, and IV. In the micrographs, dodecagonal structural units are also indicated by the white open circles. In (a) for Region I, penetrated structural units first appear along one of the $<110>_\sigma$ directions in the σ matrix. A large region characterized by a rectangular arrangement of PUI units (RAPU) then forms in (b) for Region II, where an electron diffraction pattern taken from the (σ + RAPU) coexistence state is shown in the inset. The growth of RAPU regions results in an out-of-phase domain structure, as seen in (c) for Region III. From (b) and (c), H regions are found to be nucleated from σ regions in contact with RAPU regions. The subsequent growth of H regions at the expense of RAPU regions finally leads to the approximant H state, as shown in (d) for Region IV.

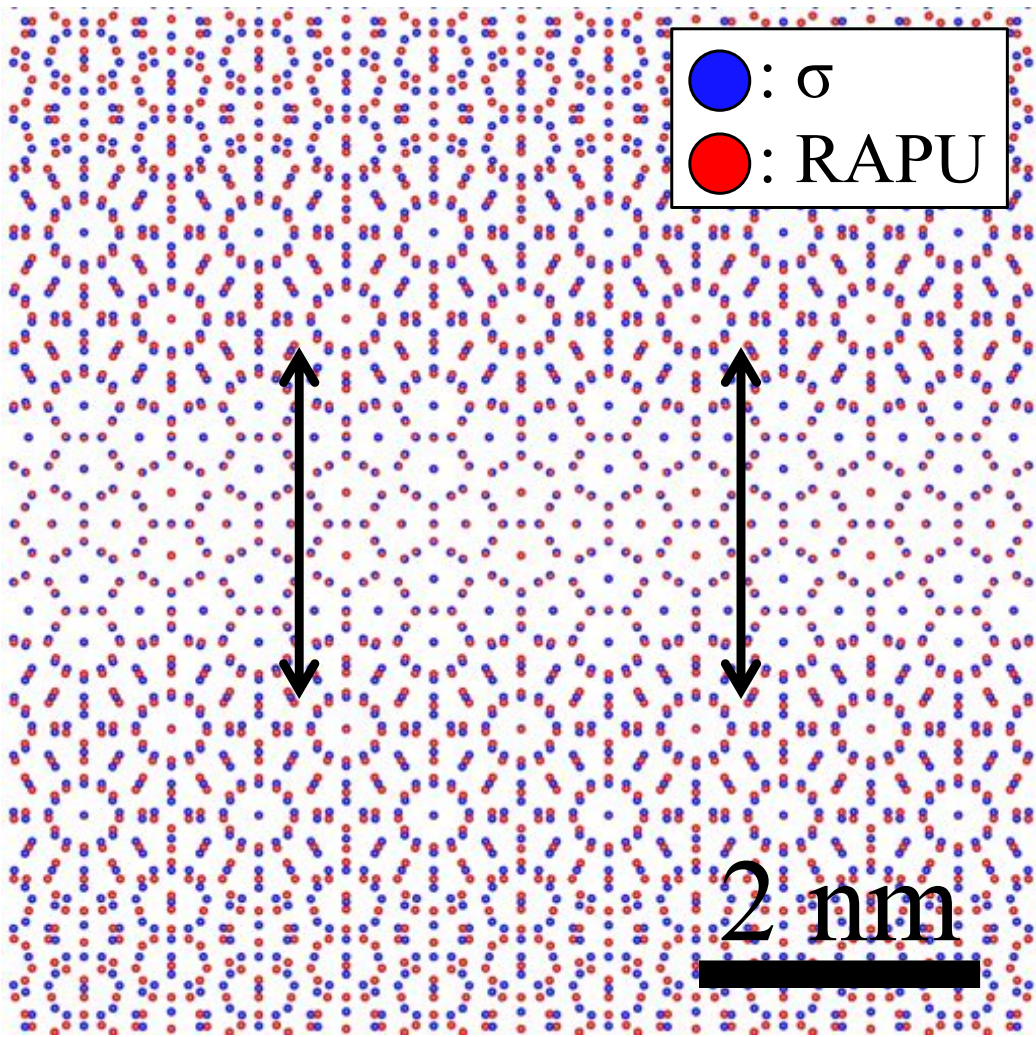

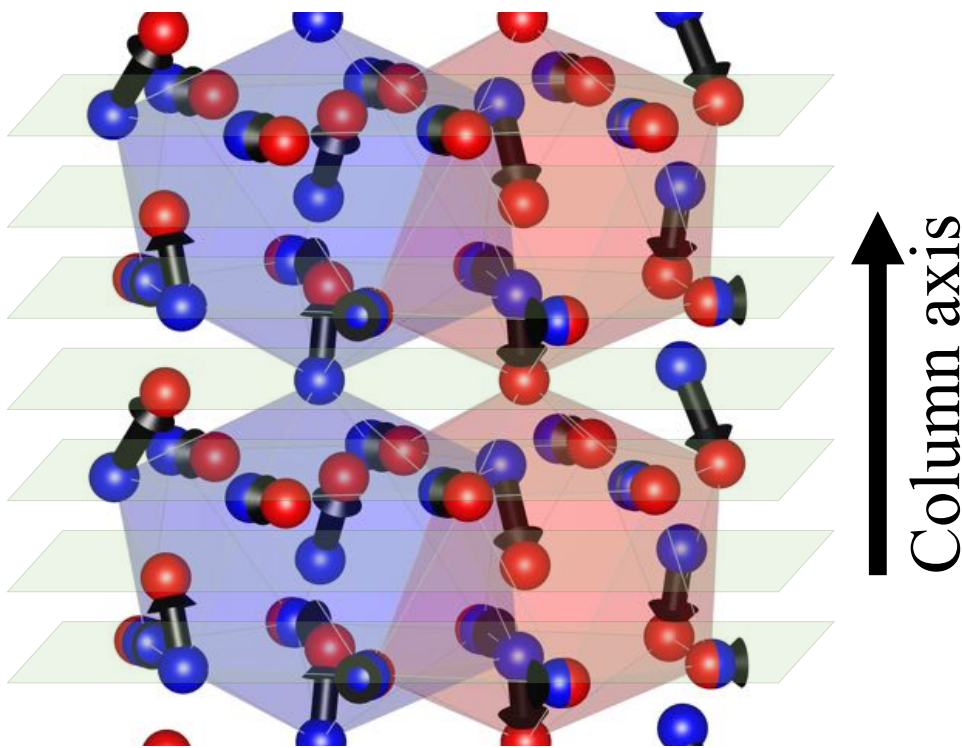

Fig. 7. $[001]_\sigma$ projection of atomic positions in both the σ and RAPU structures, together with a three-dimensional diagram for a part of the σ structure. In the diagram, atoms in the σ and RAPU states are, respectively, represented by the blue and red circles. A one-to-one correspondence between atomic positions in the σ and RAPU structures can be established in a banded region with a width of about 3 nm, as indicated by the double arrows. In the banded region, atomic positions in the RAPU structure can be obtained from those in the σ structure by simple atomic shifts mainly along the column axis $[001]_\sigma$ direction, as indicated by the thick black arrows in the three-dimensional diagram.

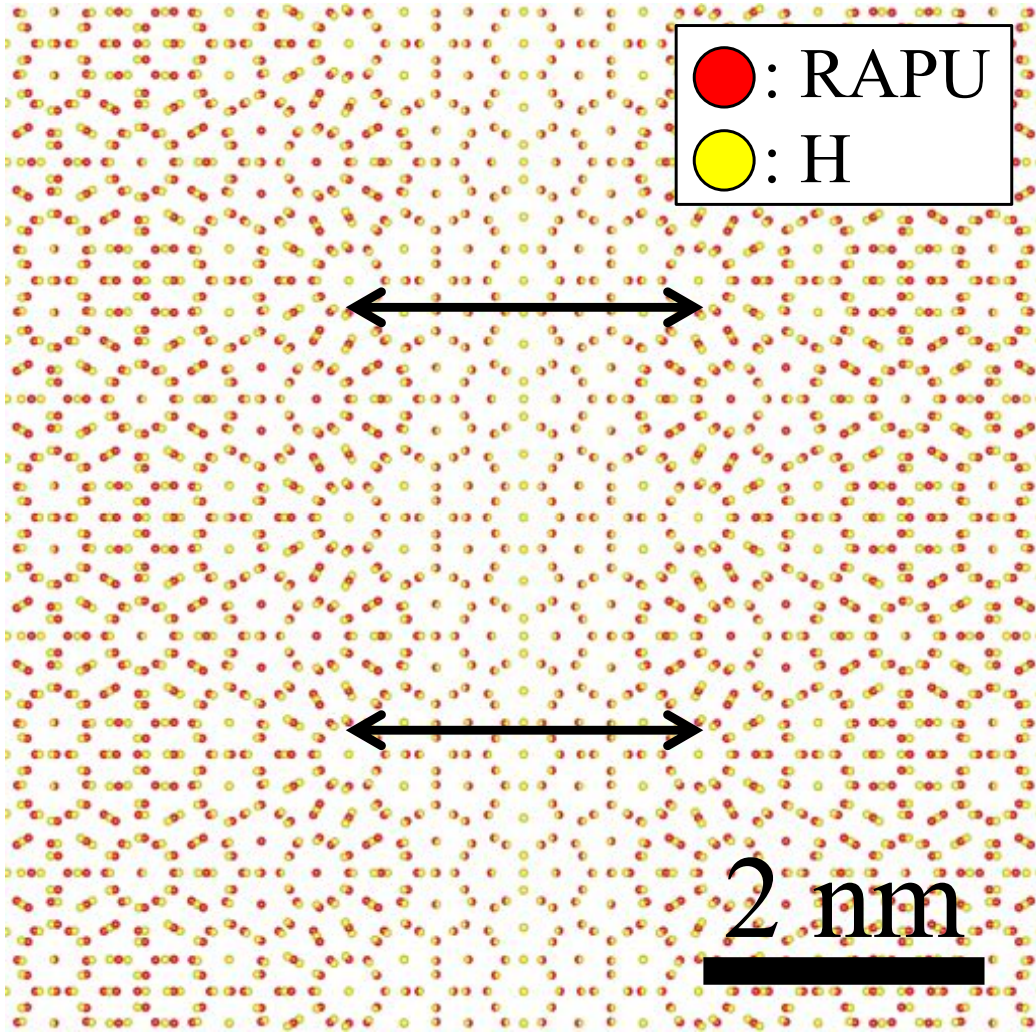

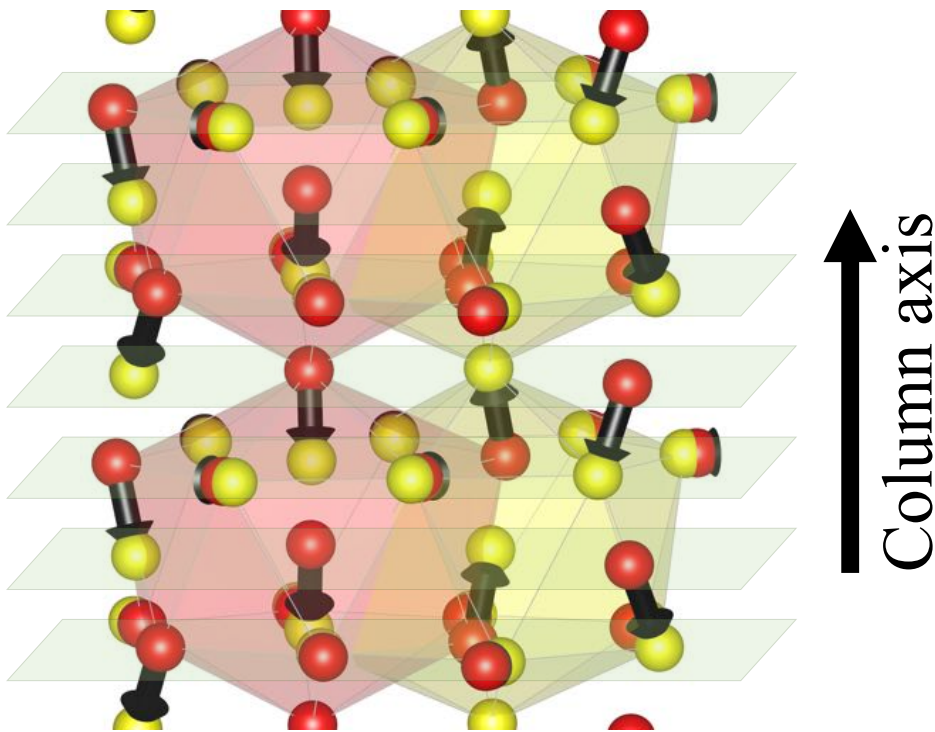

Fig. 8. Column axis $[001]_\sigma$ ($=[0001]_H$) projections of atomic positions in both the RAPU and H structures, together with a three-dimensional diagram for a part of the RAPU state. The red and yellow circles in the diagrams represent atoms in the RAPU and H structures. It is found that one-to-one correspondences are established between atomic positons of the RAPU and H structures in a banded region indicated by the double arrows in the upper diagram. As shown in the lower diagram, the (RAPU → H) conversion can occur by simple atomic shifts indicated by the black arrows, basically along the column axis direction.

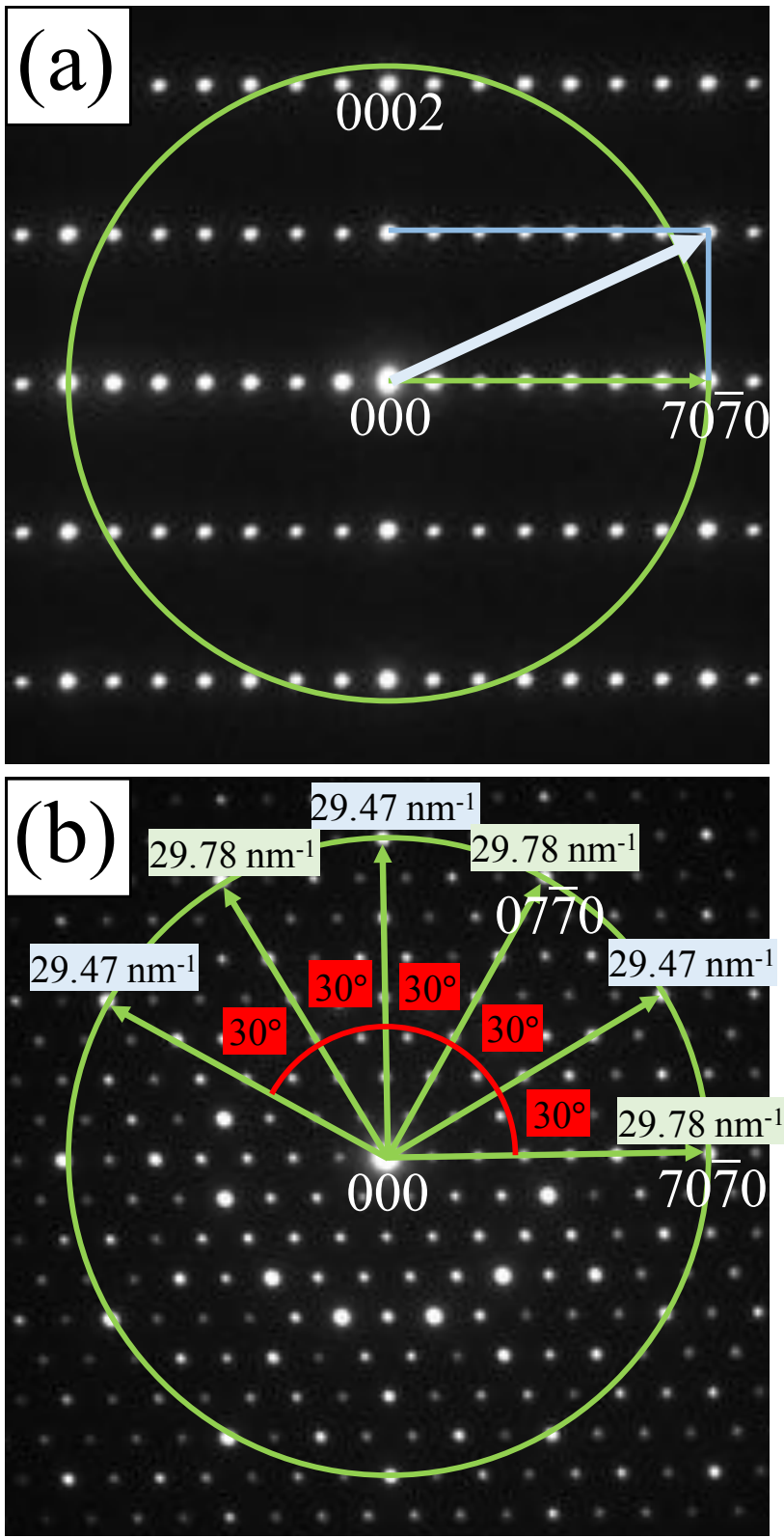

Fig. 9. Two electron diffraction patterns of the approximant H structure taken from a Mn-17at.%Si-9at.%V alloy sample annealed at 1073 K for 100 h. The electron beam incidences of the patterns in (a) and (b) are, respectively, parallel to the $[12\bar{1}0]_H$ and $[0001]_H$ directions in the H structure notation. One of the twelve wave vectors, $\mathbf{k}_j^H$ ($j$=1~12), adopted in this study is shown in (a). The adopted vector is identified as the reciprocal lattice vector for the $70\bar{7}1_H$ reflection, as is indicated by the long blue arrow. In the present study, each vector decompose into the crystallographic component of $\mathbf{k}_j^{HC} = \mathbf{k}_j^H \sin\theta$ and the pseudo-quasi-periodic component of $\mathbf{k}_j^{HP} = \mathbf{k}_j^H \cos\theta$. The latter components for the twelve waves are also indicated by the twelve green arrows in (b).

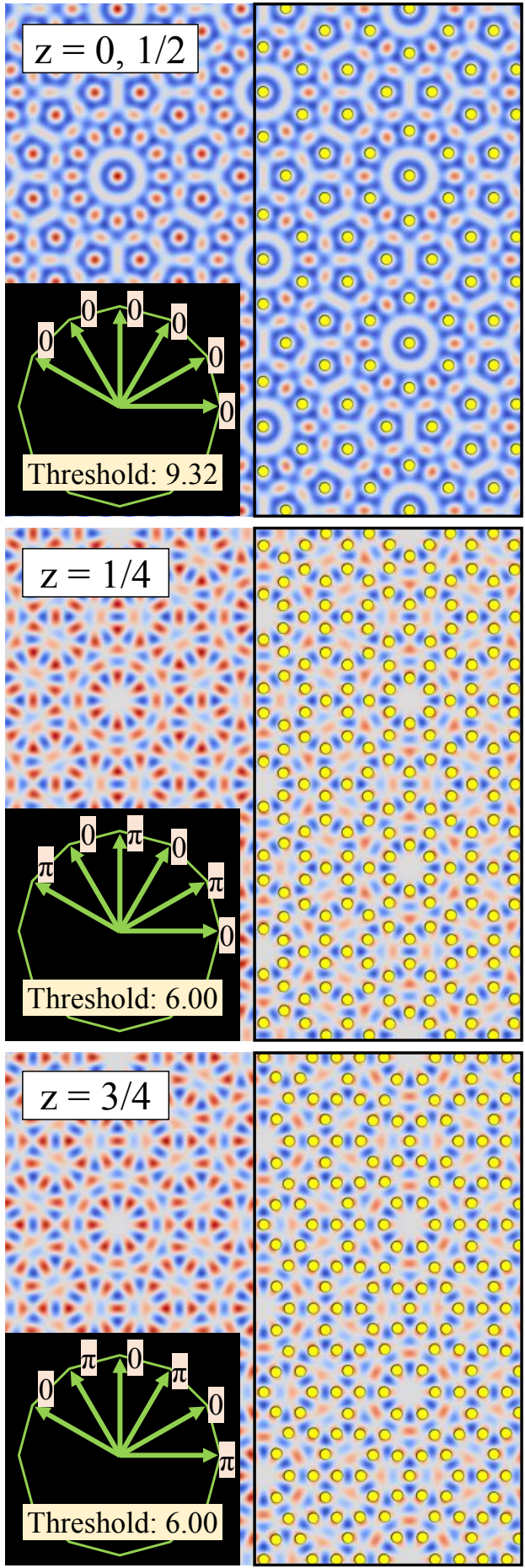

Fig. 10. Calculated distributions of $P(\mathbf{r})$ in the four layers at $z = 0$, $1/4$, $1/2$, and $3/4$ for the approximant H structure, together with the atomic positions of the H structure in the right-hand side of each distribution. The determined values of $\Delta_j$ are also shown in each inset. In the distributions, the atomic positions in the H structure are represented by the small yellow circles and those at $z = 0$ and $1/2$ are identical. The magnitude of the $P(\mathbf{r})$ value decreases in order of the orange, faint orange, faint blue, and blue colors. The calculated distributions are found to reproduce the atomic positions in the H structure by putting atoms at the positions with $P(\mathbf{r})$ values above certain thresholds. The threshold values were estimated to be 4.66 for $z = 0$ and $1/2$, 3.00 for $z = 1/4$, and 3.06 for $z = 3/4$. Atomic positions in the H structure are well reproduced with our simple plane-wave model.

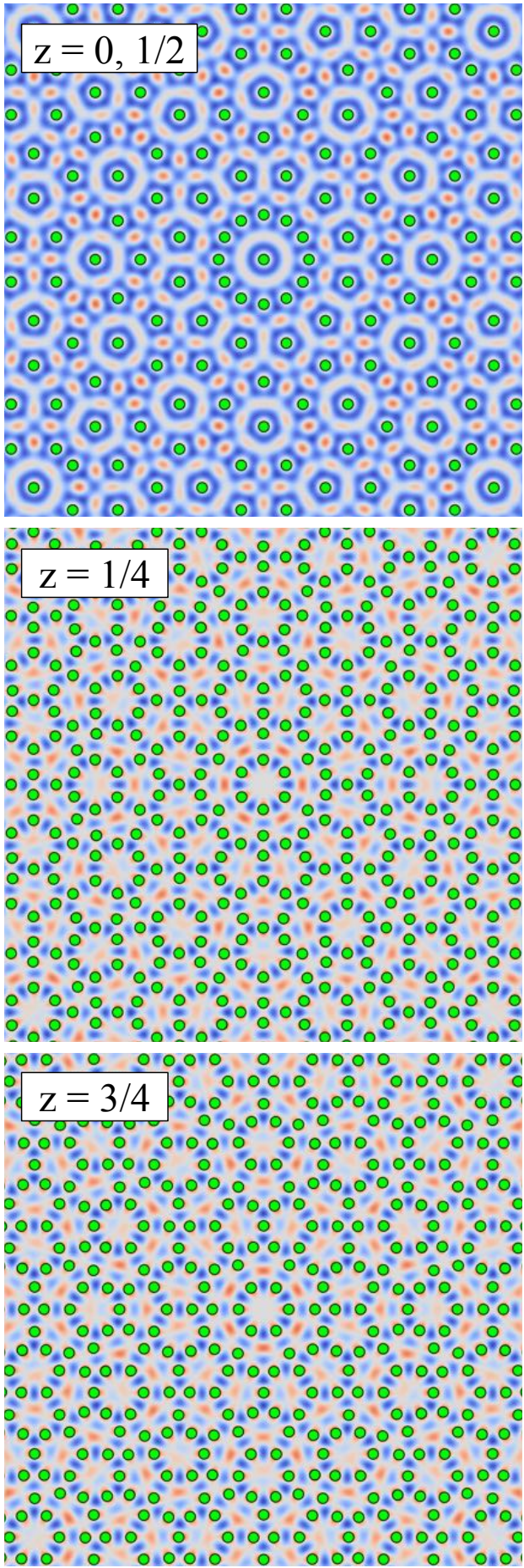

Fig. 11. Calculated distributions of $P(\bm{r})$ for the four layers at z = 0 and 1/2, 1/4, and 3/4 for the dodecagonal quasicrystal, together with its predicted atomic positions indicated by the small green circles. The atomic positions in the dodecagonal quasicrystal were predicted by assuming the same magnitude of $\bm{k}_j^{HP}$ for the twelve plane waves to obtain dodecagonal symmetry. The same $\Delta_j$ and threshold values as those obtained for the approximant H structure were also used in the calculation.

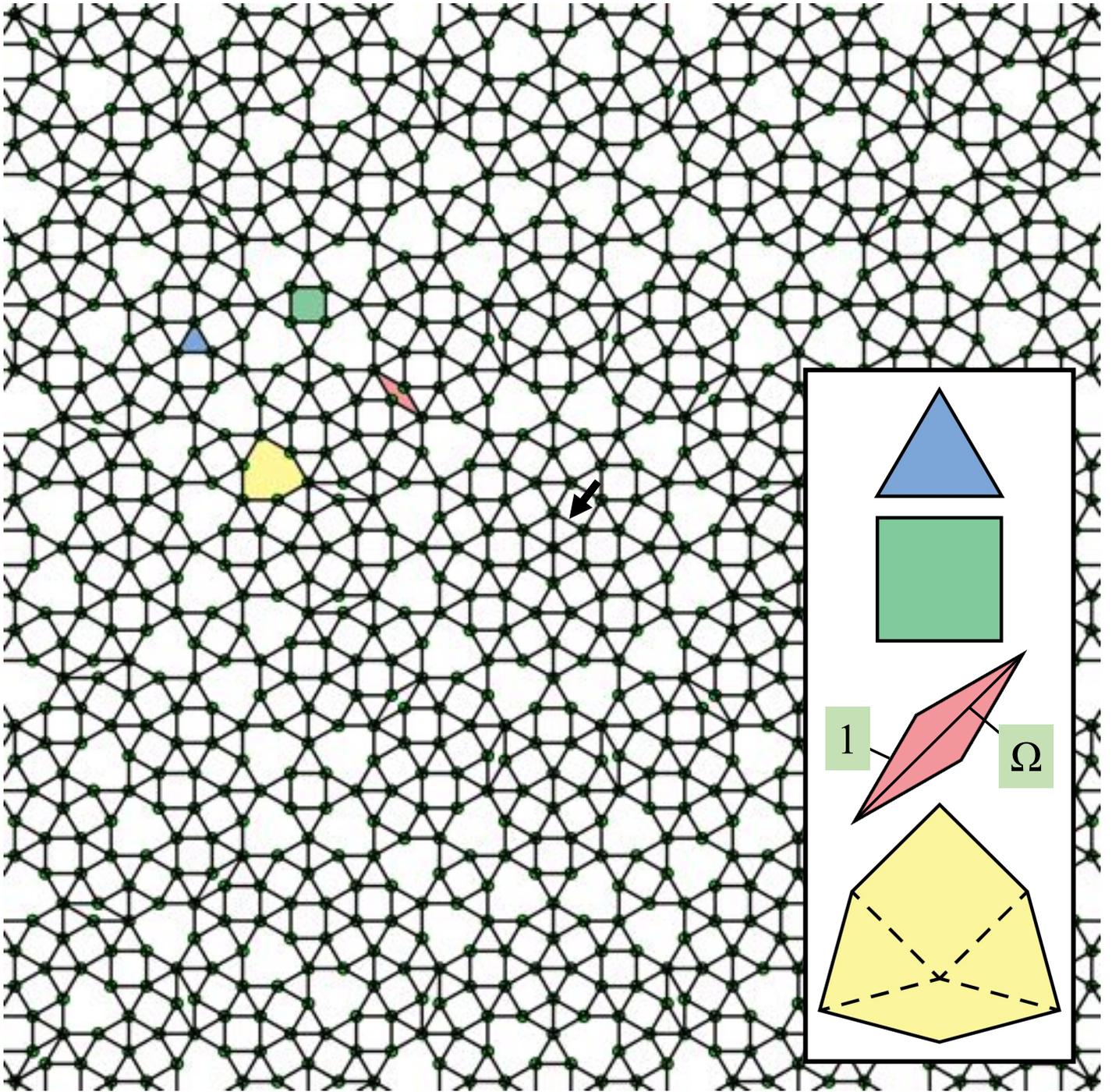

Fig. 12. Calculated tiling pattern of the dodecagonal quasicrystal predicted with our plane-wave model, together with the four types of tiles present in the calculated pattern. The tiling pattern was actually constructed by connecting nearest-neighbor atoms at distances from 0.452 nm to 0.456 nm in the layer at z = 0. Square-, triangle-, rhombus-, and distorted hexagonal-shaped tiles were found, which are indicated by the green, blue, red, and yellow colors, respectively. Because the distorted hexagonal-shaped tile is an assembly of two triangles, one square, and one rhombus, the pattern of the dodecagonal quasicrystal is thus understood to consist of triangle-, square-, and rhombus-shaped tiles.